\documentclass{cjaa}                   % preprint, the final version for
\input{epsf.sty}                        %for PS/EPS graphics inclusion, old

\newcommand{\kms}{km~s$^{-1}$}

\begin{document}

   \title{Merger Dynamics of the Pair of Galaxy Clusters --- A399 and A401
\,$^*$ \footnotetext{$*$ Supported by the National Natural Science
Foundation of China.}
%   \subtitle{I. Place Your Subtitle Here}
}
 \volnopage{Vol.0 (200x) No.0, 000--000}
%%preserved for Editor. DOn't remove!
 \setcounter{page}{1}
 %%starting page, preserved for Editor. DOn't remove!
%   \baselineskip=5mm             %%preserved for Editor. DOn't remove!

   \author{Qi-Rong Yuan \inst{1,3}, Peng-Fei Yan  \inst{2},
%      \inst{1}\mailto{yuanqirong@njnu.edu.cn}
%% Please move "\mailto{}" to the corresponding author of the paper.
%% For single author or all the authors from an institute, use "\inst{}" only
%% Here is an example of three authors come from different institutes.
    Yan-Bin Yang\inst{3}, \and Xu Zhou \inst{3}
      }
   \offprints{Q.-R. Yuan}                   %% is disabled in fact

   \institute{Department of Physics, Nanjing Normal University,
                 Nanjing 210097, China\\
             \hspace{3.5mm}{yuanqirong@njnu.edu.cn}
%% Please give the E-mail address of the author, to whom future
%% correspondence and
%% offprint requests will be sent. Note to pair \mailto{} with \email{}
        \and
        Department of Mathematics and Physics, Qingdao University of Science
and Technology, Qingdao 266042, China\\
       \and
           National Astronomical Observatories, Chinese Academy of Sciences,
             Beijing 100012, China\\
             %\email{zhouxu@vega.bac.pku.edu.cn}
          }

   \date{Received~~2004 July 15; accepted~~2005 February 12}

   \abstract{
Convincing evidence of a past interaction between two rich clusters A399 and
A401 was recently found by the X-ray imaging observations. In this paper we
examine the structure and dynamics of this pair of galaxy clusters. A
mixture-modeling algorithm has been applied to obtain a robust partition
into two clusters, which allows us to discuss the virial mass and velocity
distribution for each cluster. Assuming that these two clusters follow a
linear orbit and they have once experienced a close encounter, we model the
binary cluster as a two-body system. As a result, four gravitationally bound
solutions are obtained. The recent X-ray observations seem to favor a
scenario in which the two clusters with a true separation of
$5.4h^{-1}\,$Mpc are currently expanding at 583 \kms along the direction
with a projection angle of $67^{\circ}.5$, and they will reach a maximum
extent of $5.65\,h^{-1}$\,Mpc in about $1.0\,h^{-1}$ Gyr.
   }

   \authorrunning{Q.R. Yuan, P.F. Yan, Y.B. Yang \& X. Zhou }
%author_head in even pages
   \titlerunning{Merger Dynamics of the Cluster Pair A399/A401 }
% title_head in odd pages

   \maketitle

%% The author head (on even pages) and the title head (on odd pages) will be
%% automatically extracted from \author{} and \title{}. Whenever the title is
%% too long, you will be asked to supply a shorter one by inserting either
%% \authorrunning{} or
%% \titlerunning{} before \maketitle. Anyway, you can specify your own heads
%% in advance.
%% Note: In the following text body of your manuscript, please note several
%% differences from other major journals:
%% (1) \subsection{Please Capitalize the First Letter of Each Notional Word
%in
%% Subsection Title}
%% (2) Please Capitalize the First Letter of Each Notional Word in table's
%% caption
%
%________________________________________________ sections below
%

\section{Introduction}
%% first-level sections will be auto-capitalized
\label{sect:intro}
%\hspace{15pt}%                   %% preserved for Editor
According to the hierarchical bottom-up scenario, clusters of galaxies are
thought to be formed by accreting and merging of subunits. The structure and
dynamics of the rich galaxy clusters with ongoing merger events are of great
importance for understanding the cluster evolution. N-body numeric
simulations show that substructure may occur in individual rich clusters
before their final collapse and virialization (White 1976; Burns et al.\
1994). Since the cluster merging events are the most common and energetic
phenomena in the Universe, more and more observational efforts in optical
and X-ray bands have been devoted to the nearby rich clusters with
significant substructures (e.g., White, Briel \& Henry 1993; Gastaldello et
al.\ 2003).

When these subsystems are slightly more separated they may be classified as
separate galaxy clusters. The interacting system of clusters A399 and A401
is a good example. They have long been treated as a merging pair of clusters
since they are close to each other, with a projected separation of $\sim 0.6
^\circ$ between their central cD galaxies (McGlynn \& Fabian 1984; Oegerle
\& Hill 1994). The temperature map for this binary cluster, derived from the
$ASCA$ spatially resolved spectroscopic data, possibly suggests a physical
link or a massive dark matter filament between these two clusters
(Markevitch et al.\ 1998). The X-ray observation with the {\it ROSAT} High
Resolution Imager (HRI) unveiled a significant linear structure of A399
which points directly to the core of A401, and this feature might be
resulted from a past violent interaction (Fabian, Peres \& White 1997).
Recent {\it XMM-Newton} observations also confirmed the enhanced X-ray flux
and temperature in the region between two clusters (Sakelliou \& Ponman
2004).

Therefore it is of great interest to search for the optical anomalies in
dynamics of the member galaxies. In general, the merger history can be
modelled on the basis of the structure and dynamics of cluster galaxies,
intracluster gas, and intergalactic medium (e.g., Colless \& Dunn 1996).
This paper will use the existing redshift measurements to investigate the
possible structures in 2-dimension map and in velocity space. A prevalent
mixture-modeling algorithm, known as the KMM algorithm (Ashman et al.\
1994), will be applied to obtain a robust separation of these two clusters.
Disregarding the rotation of the system, we will try to model this cluster
pair as a two-body system on the basis of the velocity distribution and
virial mass estimates.  In \S2, we present the spatial distribution and
localized variations in velocity distribution for all the member galaxies in
the A399/A401 system as a whole. We apply the KMM partition algorithm and
discuss the velocity distribution for each cluster in \S3. Then, the virial
mass estimate and two-body modeling for this binary system of galaxy
clusters are performed in \S4. Finally, a summary is given in \S5.

%% ChJAA editors DID NOT use \cite{} for citation, \ref and \label for
%% cross-references of Table/Figure in publication version.
%% ChJAA editors prefered you giving a citation as 'Michel et al.\ 1992', and
%% writting Table~1 or Fig.~1 and so forth. However, that will make authors
%% inconvenient in adjusting/adding/removing text, tables or figures. Anyway,
%% authors can use \cite, \citep and \citet as widely used in other journals.
%% ChJAA editors are moving to use a more flexible LaTeX source.

\section{PROPERTIES OF THE SAMPLE}

A399 and A401 are nearby clusters of galaxies ($z \sim$ 0.07181 and 0.07366;
Oegerle \& Hill 2001), classified respectively as type I-II and I clusters
by Bautz \& Morgan (1970). With respect to the center of this binary system
($2^{\rm h}58^{\rm m}30^{\rm s}, +13^{\circ}20''$; J2000.0), 1331
extragalactic objects with positional offsets less than 100.0 arcmin were
extracted from the NASA/IPAC Extragalactic Database (NED). However, only 240
galaxies have spectroscopic redshifts. The remainder appear only in one of
the imaging surveys from radio, infrared, and X-ray wavebands.

Most of the spectroscopic data were contributed by Hill \& Oegerle (1993)
who carried out a survey of the cD clusters. The redshift measurements were
detailed in Hill \& Oegerle (1993), and the typical velocity uncertainty for
the galaxies is less than 100 \kms. The distribution of spectroscopic
redshifts for 240 known galaxies is shown in Fig.~1. There are 215 galaxies
with their redshifts in a range of 18,000 \kms $< cz <$ 25,000 \kms, with a
peak at $\sim$ 21,500 \kms. Only one peak can be found in the velocity
distribution because the velocity dispersions for individual clusters are
larger than the apparent velocity separation between two clusters. It is
unambiguous to treat these 215 galaxies as the members of this cluster pair
since their redshift distribution is quite concentrated and isolated. The
contamination from the foreground and background galaxies should be very
slight. The spatial distribution for these 215 galaxies is presented in
Fig.~2. We superpose the contour map of the surface density that has been
smoothed by a Gaussian window with $\sigma = 2'$. Because of the severe
overlap in the redshift distributions, clusters A399 and A401 can not be
completely resolved by the velocity distribution only.

%and the dynamics of this merger system has been sketchily discussed by
% Oegerle \& Hill (1994).

\begin{figure}[b]
\begin{center}
%\epsscale{0.4}{0.4} \plotone{fig1.eps}
\mbox{\epsfxsize=0.7\textwidth\epsfysize=0.65\textwidth\epsfbox{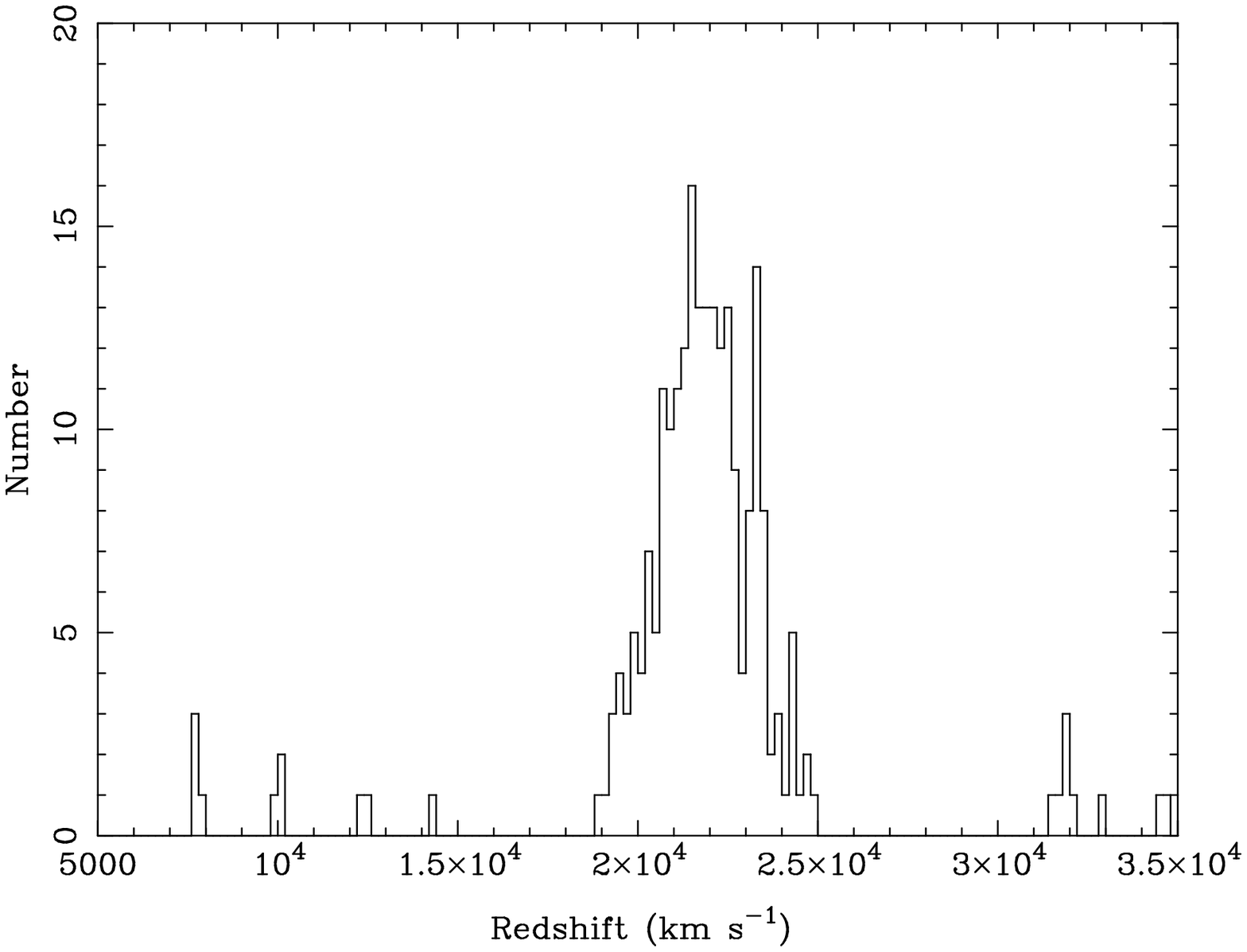}}
\caption{Distribution of the radial velocities for 240 galaxies
with known spectroscopic redshifts. The bin size is 1000 \kms }
\label{Fig:plot1}
\end{center}
\end{figure}

\begin{figure}[h]
\begin{center}
%\epsscale{0.4}{0.4} \plotone{fig2.eps}
\mbox{\epsfxsize=0.6\textwidth\epsfysize=0.6\textwidth\epsfbox{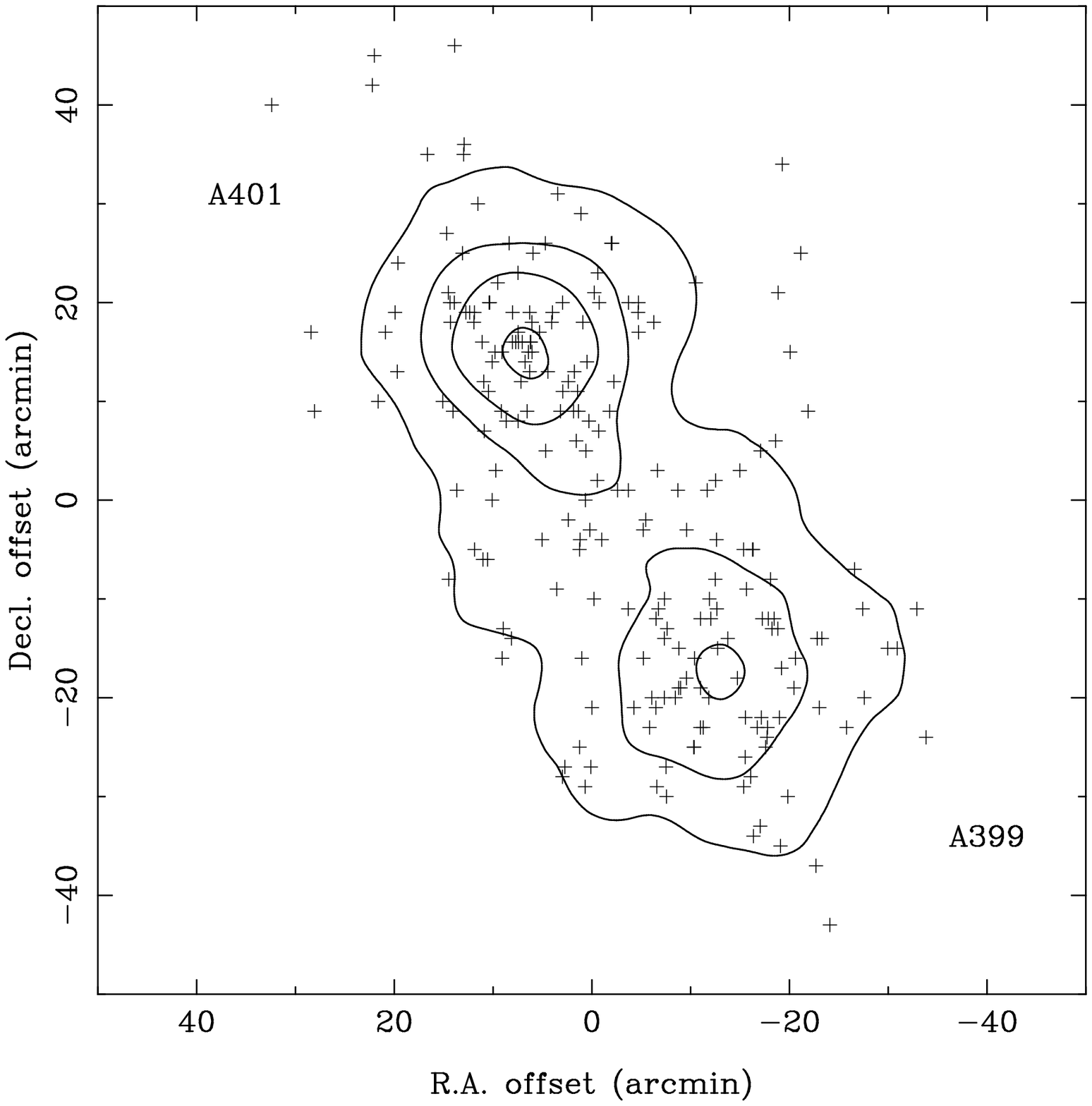}}
\caption{Spatial distribution for 215 member galaxies of the
binary system of galaxy clusters, superposed by the contour map of
the surface density where the smoothing Gaussian window with a
radius of $2'$ is used. The contour levels are 0.03, 0.09, 0.15,
and 0.21 arcmin$^{-2}$, respectively. } \label{Fig:plot2}
\end{center}
\end{figure}

In order to detect the substructures in both the velocity space
and the projected map, we take use of the $\kappa$-test (Colless
\& Dunn 1996) for the A399/A401 system as a whole. The statistic
${\kappa}_n$ is defined to characterize the local deviation on the
scale of groups of $n$ nearest neighbors. A larger ${\kappa}_n$
means a greater possibility that the local velocity distribution
differs from the overall distribution. The probability that
$\kappa_n$ is larger than the observed value $\kappa_n^{\rm obs}$,
$P(\kappa_n>\kappa_n^{\rm obs})$, can be estimated by Monte Carlo
simulations by randomly shuffling velocities. Table 1 gives the
results of $\kappa$-test for 215 known member galaxies, and $10^3$
simulations are used to estimate $P(\kappa_n>\kappa_n^{\rm obs})$
for all cases.

It is found that the optimum scale of the nearest neighbors is 10,
and the substructure appears most obvious on this scale. The
bubble plot in Fig.~3 shows the location of localized variation
using neighbor size $n=10$, and the bubble size for each galaxy is
proportional to $-\log[P_{\rm KS}(D>D_{\rm obs})]$. Therefore
larger bubbles indicate a greater difference between local and
overall velocity distributions.  A comparison with Fig.~2 shows
that the bubble clusterings near the central regions of A399 and
A401 are found to be significant, which indicates a distinct
difference between the localized and whole velocity distributions.

%table 1
\begin{table}
\caption[]{Result of $\kappa$-Test for 215 Member Galaxies in the Binary
System}
%\vspace{-9mm}
\begin{center}
\begin{tabular}{cccccccccc}   \hline
\noalign{\smallskip}
 $n$ & 4 & 5& 6& 7& 8& 9& 10&11 &12\\
\hline \noalign{\smallskip}
$P(\kappa_n>\kappa_n^{\rm obs})$ & 18.3\%&
44.4\%& 42.6\%& 34.1\%& 17.0\%& 11.9\%& 6.8\% & 15.1\%& 14.5\%\\
\noalign{\smallskip}   \hline
\end{tabular} \end{center}
\end{table}

\begin{figure} \begin{center}
%\epsscale{1.0} \plotone{fig3.eps}
\mbox{\epsfxsize=0.6\textwidth\epsfysize=0.6\textwidth\epsfbox{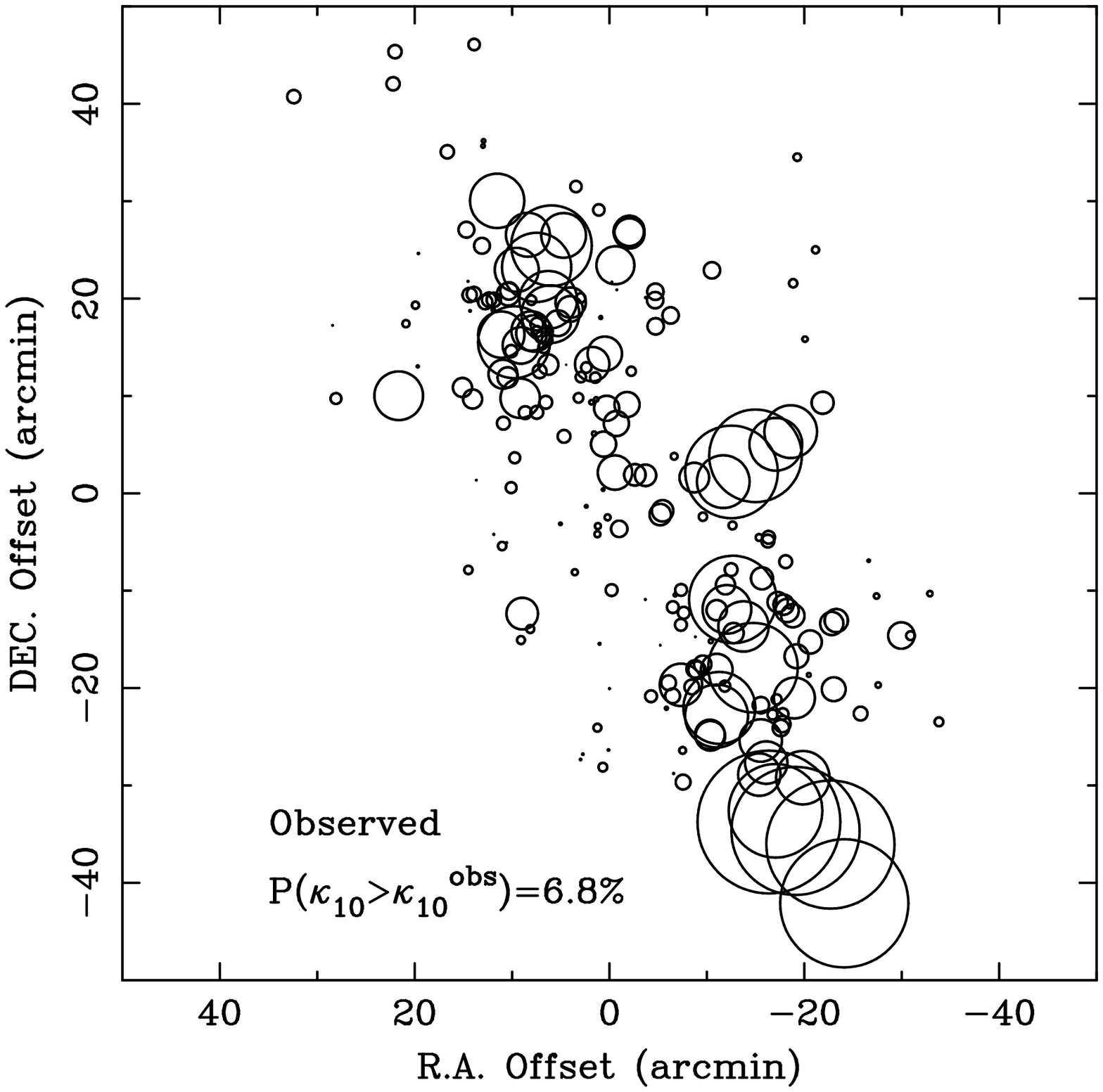}}
\caption{ Bubble plot showing the degree of difference between the
local velocity distribution for groups of ten nearest neighbors
and the overall distribution of 215 known cluster galaxies. }
\end{center} \end{figure}

\section{THE KMM PARTICIPATION INTO TWO CLUSTERS }

For studying the dynamical properties for each cluster, those 215 galaxies
should be correctly partitioned. It is relatively easy to partition the
galaxies whose projected locations appear close to the cluster centers.
However, for the galaxies located exactly between the cluster centers, the
partition might become a rather ambiguous task.

In order to achieve a robust partition, we apply a prevalent technique of
mixture modeling, namely KMM algorithm, on the sample of 215 galaxies. The
KMM is a maximum-likelihood algorithm which assigns objects into groups and
assesses the improvement in fitting a multi-group model over a single group
model (Ashman et al.\ 1994).  A detailed description of the KMM algorithm
can be also found in Nemec \& Nemec (1993).  For a dynamically relaxed
cluster, the line-of-sight velocities of galaxies are expected to be
Gaussian distributed.  Since A399 and A401 are two gravitationally distinct
clusters of galaxies, we here apply the KMM algorithm which estimates the
statistical significance of bimodality based on the three-dimension data:
the projected positions of the galaxies and the radial velocity, just as
Colless \& Dunn (1996) did. When an initial partition into two clusters or a
set of initial parameters for each cluster is given, the KMM algorithm can
start iterating toward the maximum-likelihood solution.  Table 2 lists the
various initial partitions/parameters that we used and the corresponding
final solutions, where $(\bar{x}_1, \bar{y}_1, \bar{v}_1)$ and $(\bar{x}_2,
\bar{y}_2, \bar{v}_2)$  are the mean positions and velocities of A401 and
A399, respectively, $(\sigma_{x_1}, \sigma_{y_1}, \sigma_{v_1})$ and
$(\sigma_{x_2}, \sigma_{y_2}, \sigma_{v_2})$ are their standard deviations
in positions and velocities, and $f_1$ and $f_2$ are the fractions of the
sample in the two clusters. Furthermore, the estimate of the overall
correction rate ($R$) given by the KMM algorithm is listed.

%table 2
\begin{table}[t]
\caption[]{Initial parameters and final solutions of the KMM algorithm}
\vspace{-9mm}
\begin{center}
\begin{tabular}{ccccccc}   \hline
\noalign{\smallskip}
 Case & $(\bar{x}_1, \bar{y}_1, \bar{v}_1)$ & $(\sigma_{x_1}, \sigma_{y_1},
\sigma_{v_1})$ & $(\bar{x}_2, \bar{y}_2, \bar{v}_2)$ & $(\sigma_{x_2},
\sigma_{y_2}, \sigma_{v_2})$ & $(f_1, f_2)$ & Rate \\
\noalign{\smallskip} \hline \noalign{\smallskip}
 & & &Initial Parameters & & & \\
1& (5.9,16.9,22133) & (9.8,9.9,1208) & (-11.4,-15.7,21536) & (9.8,9.5,1227) &
(0.521,0.479) & \\
2& (4.0,14.6,22080) & (10.7,11.1,1212) & (-11.7,-18.3,21505) &
(10.3,7.7,1234)& (0.595,0.405) & \\
3& (5.0,16.8,22126) & (10.5,9.7,1204) &
(-9.0,-18.1,21378) & (10.4,8.9,1232) & (0.530,0.470) &  \\
\noalign{\smallskip}   \hline \noalign{\smallskip}
 & & & Final Solutions & & & \\
1 & (4.8,14.4,22107) & (10.2,11.7,1185) & (-12.6,-17.4,21477) &
 (9.1,8.8,1241) & (0.586,0.414) & 95.2\% \\
2 & (4.9,14.5,22107) & (10.2,11.7,1185) & (-12.6,-17.3,21477) &
 (9.2,8.9,1240) & (0.585,0.415) & 95.2\%\\
3 & (4.8,14.4,22107) & (10.2,11.7,1185) & (-12.6,-17.4,21477) &
 (9.1,8.8,1241) & (0.587,0.413) & 95.1\% \\
\noalign{\smallskip}   \hline
\end{tabular} \end{center}
\end{table}

We chose to specify initial positions and dispersions for two clusters in
case 1, while in case 2 we specify a simple partition of sample in which all
galaxies with the declination offset $y>-5$ arcmin are assigned to be A401
members. With the different initial parameters we specified,  the KMM
algorithm searched for an optimum two-group solution, and converge to a very
similar results. The estimate of the correct allocation rate reaches 95\%.
In case 3 we tried another initial partition: A399 concentration includes
all galaxies within an angular distance of 20 armin to the central cD galaxy
UGC~2438, and we obtain the same final solution.

According to the final partition into two clusters, there are 127 galaxies
belonging to A401, and 88 galaxies for A399. The spatial distribution for
each cluster is plotted in Fig.~4(a). Then, we apply the $\kappa$-tests
again for individual clusters,  and the probability
$P(\kappa_n>\kappa_n^{\rm obs})$ is estimated for each cluster. Table 3
presents the $\kappa$-test results, and the corresponding bubble plot are
given in Fig.~4(b). Compared with the last $\kappa$-test on the whole binary
system (see Table 1 and Fig.~3), the variation of localized velocity
distribution for each cluster is significantly decreased, which indicates
that the KMM algorithm arrived at a robust partition.

%\clearpage
\begin{figure}[t]
%\epsscale{1.0}
\plottwo{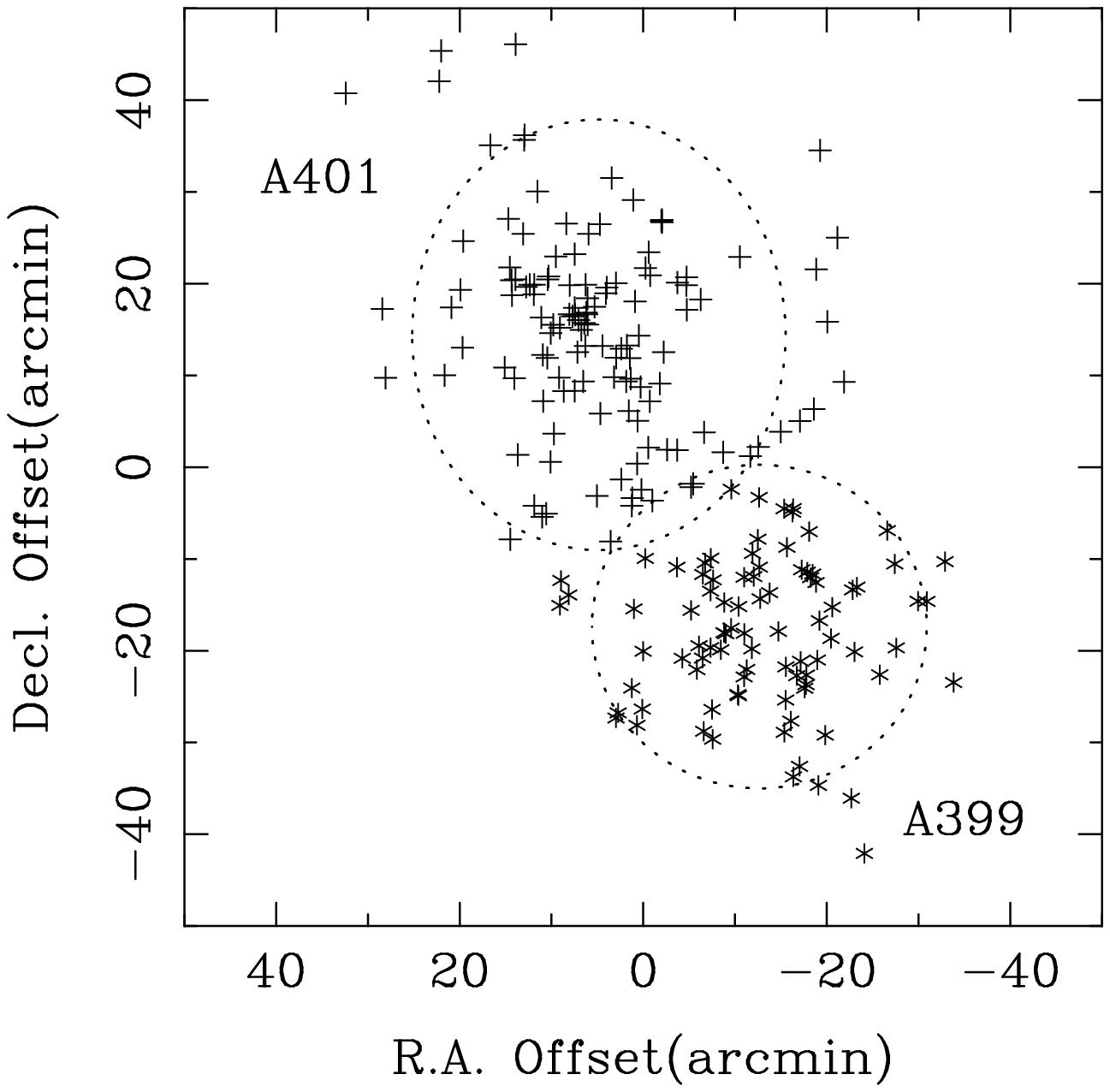} {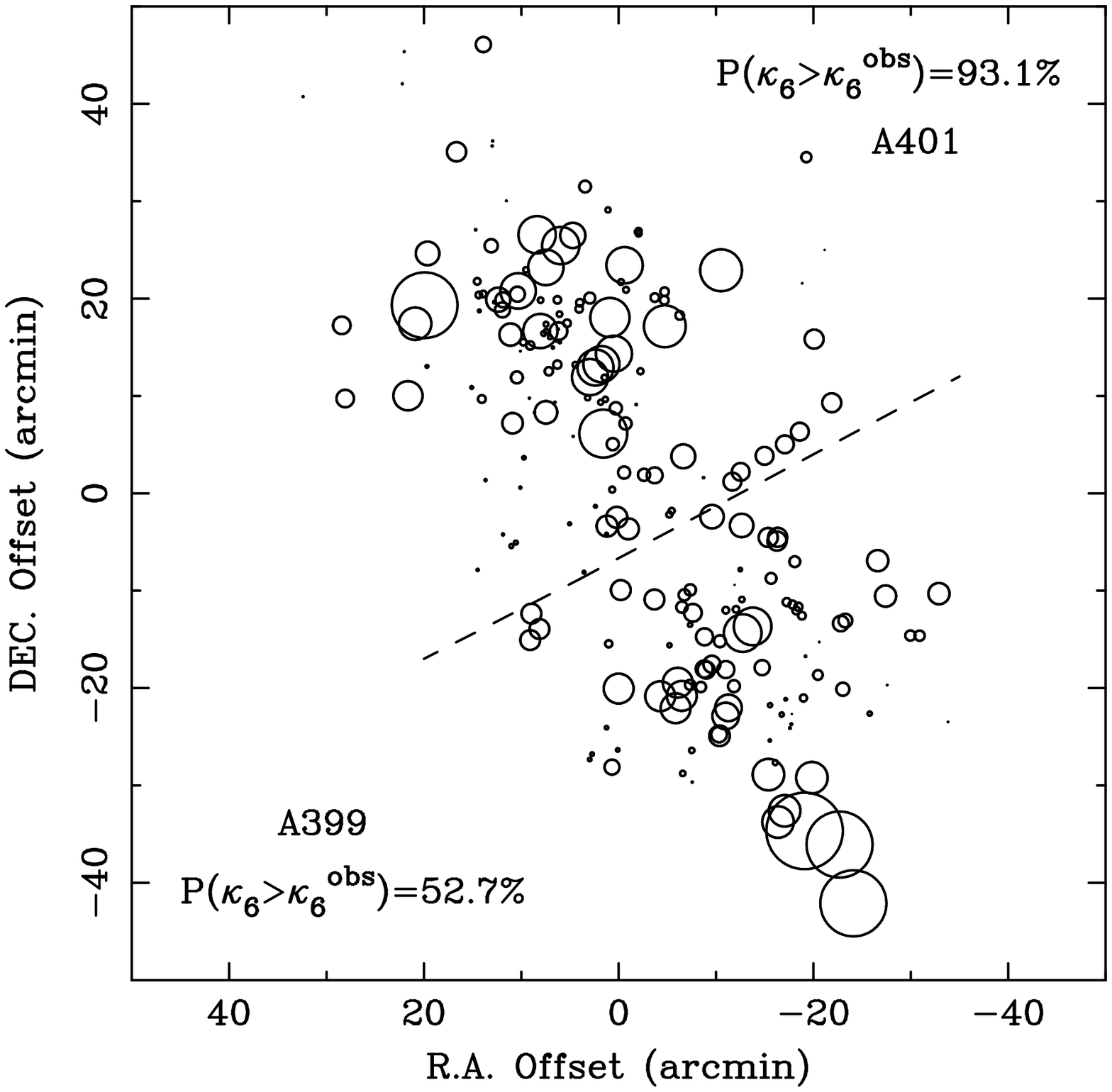} \caption{(a) The projected
positions for the member galaxies of A399 (denoted by astroids)
and of A401 (denoted by plus sign). The dotted ellipses are the
$2\sigma$ contours of the fitted Gaussians. (b) Bubble plots for
groups of six nearest neighbors for 127 galaxies in A401 and 88
galaxies in A399. The dashed line separates two clusters.}
\end{figure}

%table 3
\begin{table}[h]
\caption[]{Results of $\kappa$-tests for 88 galaxies in A399 and for 127
galaxies in A401} \vspace{-5mm}
\begin{center}
\begin{tabular}{cccccccc}
\noalign{\smallskip} \hline
Group size $n$ & 2& 3 & 4 & 5& 6& 7& 8\\
\noalign{\smallskip} \hline  $P(\kappa_n>\kappa_n^{\rm obs})$ for A399
 &33.5\%
&8.7\% & 17.5\%&
56.7\%& 52.7\%& 57.1\%& 59.1\%\\
$P(\kappa_n>\kappa_n^{\rm obs})$ for A401 &7.9\% & 59.4\%& 59.0\%&
87.3\%& 93.1\%& 87.6\%& 72.0\%\\
\noalign{\smallskip} \hline
\end{tabular} \end{center}
\end{table}

\section{MERGER DYNAMICS BETWEEN A399 AND A401}

\subsection{Velocity Structure}

The radial velocity distributions for the binary system and each cluster are
given in Fig.~5.  The solid lines represent the best-fit Gaussians with the
mean velocities and dispersions listed in Table 2. To characterize the
kinematical properties of clusters of galaxies, two robust estimators
analogous to the velocity mean and standard deviation, namely the biweight
location ($C_{BI}$) and scale ($S_{BI}$), are defined by Beers et al.\
(1990). These two quantities are resistant in the presence of outliers, and
robust for a broad range of probable non-Gaussian underlying populations.
Using the $ROSTAT$ software, we compute the biweight location and scale in
line-of-sight velocity distribution for each cluster. As a result, we obtain
$C_{BI}=21491\pm141$ \kms\ and $S_{BI}=1315 \pm 82$ \kms\ for A399, and
$C_{BI}=22069\pm107$ \kms\ and $S_{BI}=1212 \pm 71$ \kms\ for A401.  The
physical line-of-sight velocity difference between these two clusters is
$V_r=\Delta (C_{BI})/(1+\bar{z})=539\pm165$ \kms.

%\clearpage
\begin{figure}[t] \begin{center}
\mbox{\epsfxsize=0.6\textwidth\epsfysize=0.9\textwidth\epsfbox{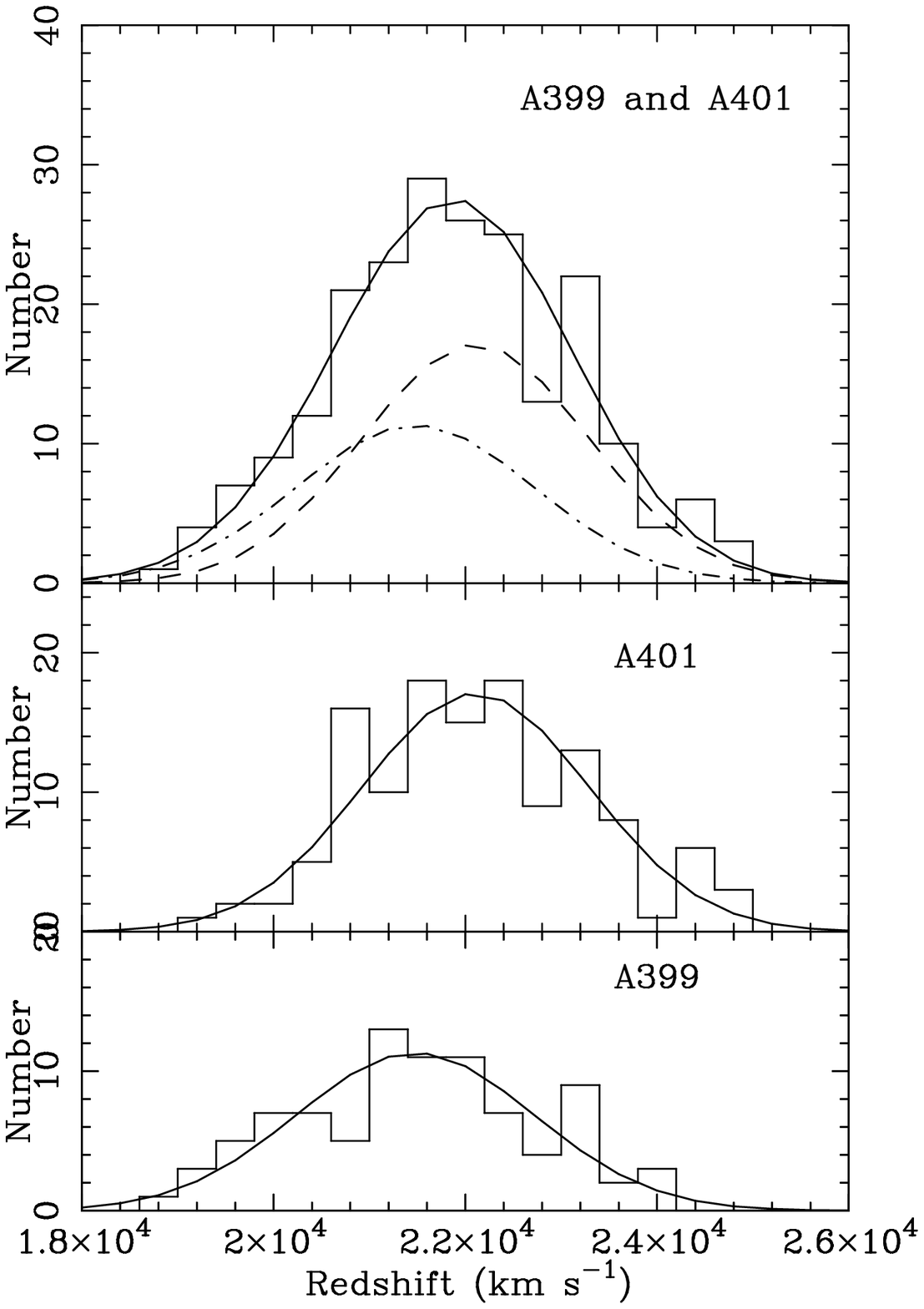}}
%\epsscale{0.4}{0.4}
%\plotone{fig5.eps}
\caption{The velocity distributions for the
galaxies in (a) the entire binary system, (b) A401, and (c) A399, superposed
with the fitted Gaussian distributions.} \end{center} \end{figure}

\subsection{Virial Mass Estimate}

A399 and A401 are gravitationally bound with each other. In order to
determine the dynamical state of the binary system, we will apply the virial
theorem for estimating the mass of each cluster. Assuming that the galaxy
cluster is bound and the galaxy orbits are random, the virial mass ($M_{\rm
vt}$) can be estimated from the following standard formula (Geller \&
Peebles 1973; Oegerle \& Hill 1994):
$$M_{\rm vt}=\frac{3\pi}{G}\sigma^2_{\rm r} D N_{\rm p} \left( \sum_{i>j}^{N}
\frac{1}{\theta_{ij}} \right) ,$$
where $\sigma_{\rm r}$ is the
line-of-sight velocity dispersion, $D$ is the cosmological distance of the
cluster, $N_{\rm p}=N(N-1)/2$ is the number of galaxy pairs, and
$\theta_{ij}$ is the angular separation between galaxies $i$ and $j$. The
extended X-ray emission associated with A399 and A401, revealed by the {\it
ROSAT} HRI imaging observations (Fabian et al.\ 1997), indicates that at
least the first of these assumptions is reasonable. We derive the virial
masses of $2.0h^{-1} \times 10^{15} M_{\odot}$ and $2.1 h^{-1}\times 10^{15}
M_{\odot}$ for A399 and A401, respectively.

\subsection{Two-Body Models}

With the estimate of the mass for each cluster, we can investigate the state
of evolution. The concise dynamical model for this binary system is the
two-body model which was first applied to clusters by Beers, Geller \&
Huchra (1982). In this model two clusters are treated as point masses
following a linear orbit under their mutual gravity. They are presumed to
have started with zero separation and then to have moved apart before
turning around. Given the projected separation of this binary system $R_{\rm
p}$, the relative radial velocity $V_{\rm r}$ and the total mass $M$, the
model can speculate the projection angle $\alpha$ (the angle between the
line joining the two clusters and the plane of the sky), the true separation
$R$, and relative velocity $V$. The equations of motion are as follows:
\begin{eqnarray}
V &=& \frac{V_{\rm r}}{\sin\alpha} = \left( \frac{2GM}{R_{\rm m}}
 \right)^{1/2}\, \frac{\sin
\chi}{(1- \cos \chi)}, \\
R &=& \frac{R_{\rm  p}}{\cos\alpha} = \frac{R_{\rm m}}{2}(1-\cos \chi),\\
t &=& t_0 = \left( \frac{R_{\rm m}^3}{8GM} \right) ^{1/2} (\chi-\sin \chi),
\end{eqnarray}
where $R_m$ is the maximum expansion, $M$ is the total mass of the binary
system, $t_0$ is the age of the universe, and $\chi$ is the developmental
angle tracing the merger process. The two clusters have zero separation when
$\chi=0,\,2\pi$, while they are at maximum expansion when $\chi=\pi,\,
3\pi$. Due to the ambiguity in observing the system only in projection, this
model usually results in more than one orbital solution.

The KMM analysis provides the initial estimate of the projected separation
$R_{\rm p}$ of the two-body model.  Assuming a Friedmann-Robertson-Walker
cosmology with $\Omega_m=0.3$, $\Omega_{\Lambda}=0.7$, we adopt the age of
the universe as $t_0=9.43 h^{-1}$\,Gyr $= 2.976\,h^{-1} \times 10^{17} {\rm
s}$, and the angular separation between the centroids of these two clusters
($\sim 36.3$ arcmin) corresponds to a projection distance $R_{\rm p}$ of
$2.05 \,h^{-1}$Mpc at the average redshift. We take the total mass $M = \sum
M_{\rm vt} = 4.1\,h^{-1} \times 10^{15} M_{\odot}$ in our modeling.

Previous applications of the two-body models tried to solve the dynamical
solutions within a range of $0<\chi<2\pi$, assuming that the subclusters
start to saperate at t=0 and they are moving apart or coming together for
the first time in their history (Beers, Geller \& Huchra 1982; Oegerle \&
Hill 1994; Colless \& Dunn 1996). However, numerical simulations of cluster
collisions by McGlynn \& Fabian (1984) showed that one clusters can even
pass through each other without destorying the optical components. For this
pair of clusters, recent observational evidences in X-ray and radio bands
support such a picture that A399 and A400 have been passed through each
other in the past. As mentioned in \S1, Fabian et al.\ (1997) found a linear
X-ray structure in A399 pointing at the cD galaxy of A401, indicating a past
violent interaction. On the other hand, the absence of a cooling flows in
both A399 and A401 was first found by Edge et al.\ (1992), and confirmed by
Markevitch et al.\ (1998) and Sakelliou \& Ponman (2004). According to the
numerical experiments of McGlynn \& Fabian (1984), the structure of a
cooling flow can be disrupted by the merger of two similar clusters. A
simulations by Burns et al.\ (1997) showed that mergers of clusters with a
mass ratio of 1:4 may destroy a pre-existing cooling flow. The picture of a
previous interaction between these clusters is also supported by the
temperature map for this system which shows a bridge of matter between
clusters (Markevitch et al.\ 1998). Furthermore, the extensive cluster radio
helo is found to be associated with A401, suggesting the coalescence of
clusters (Harris et al.\ 1980).

Based on above observations, we suppose that the two clusters started at
$t=0$ with zero separation, and expanded to a maximum extent, and once
experienced a close encounter, which indicates a $\chi$ range from 2$\pi$ to
4$\pi$. From the equations of motion given above, four two-body models are
allowed within $2\pi<\chi<4\pi$, including two expanding (outgoing) and two
collapsing (incoming) models.  Fig.~6 shows the ($\alpha, V_{\rm r}$)-plane
where four solutions at $V_{\rm r}=539\pm165$ \kms\ are found in the bound
region. Fig.~6 also shows the limit curve for the bound region, which can be
defined by the Newtonian criterion $V_r^2 R_{\rm p} \leq 2GM\, \sin^2\alpha
\, \cos \alpha$. Meanwhile, we derive three solutions within $0<\chi<2\pi$
for reference only. The seven solutions are presented in Table 4.

%table 4
\begin{table}[h]
\caption[]{Gravitationally bound solutions for the two-body model}
\vspace{-5mm}
\begin{center}
\begin{tabular}{ccccccc}
\noalign{\smallskip} \hline
Case & Dynamical & $\chi$ & $\alpha$ & $V$ & $R$ & $R_{\rm m}$ \\
  & Status &($\pi$ rad.) &  (degree) & (\kms) &($h^{-1}$ Mpc) & $(h^{-1}$
 Mpc)
\\
\noalign{\smallskip} \hline
(a)& Outgoing & $ 0.776$ & $81.6^{+0.6}_{-0.6}$ & $545^{+165}_{-166}$ &
 $14.09^{+1.04}_{-1.00}$ & $15.99^{+3.32}_{-2.16}$ \\
(b)& Incoming & $ 1.175$ &$75.8^{+1.2}_{-1.3}$ & $556^{+175}_{-172}$ &
 $8.38^{+0.75}_{-0.71}$ & $9.04^{+0.45}_{-0.46}$ \\
(c)& Incoming & $ 1.636$ &$9.0^{+2.8}_{-2.8}$ & $3460^{+27}_{-15}$ &
 $2.08^{+0.01}_{-0.02}$ & $7.10^{+0.00}_{-0.01}$ \\
\noalign{\smallskip} \hline \noalign{\smallskip}
(d)&Outgoing  & $ 2.374$ & $9.0^{+2.9}_{-2.8}$ & $3430^{+31}_{-22}$ &
 $2.08^{+0.02}_{-0.02}$ &$6.75^{+0.00}_{-0.01}$ \\
(e)&Outgoing  & $ 2.854$ &$67.5^{+0.2}_{-0.4}$ &$583^{+181}_{-179}$
 &$5.36^{+0.04}_{-0.09}$ &$5.65^{+0.12}_{-0.11}$ \\
(f)&Incoming  & $ 3.141$ &$64.4^{+1.1}_{-1.4}$ &$598^{+192}_{-187}$ &
 $4.74^{+0.21}_{-0.23}$ & $4.98^{+0.09}_{-0.07}$ \\
(g)&Incoming  & $ 3.523$ &$10.3^{+3.4}_{-3.2}$ &$3008^{+33}_{-35}$
 &$2.08^{+0.03}_{-0.01}$ &$4.48^{+0.01}_{-0.00}$ \\
\noalign{\smallskip} \hline
\end{tabular} \end{center}
\end{table}

Fig.~6 shows that A399 and A401 are very likely to be gravitationally bound
unless the projection angle $\alpha$ is smaller than $7^{\circ}$. The
unbound orbit requires the true relative velocity $V > 4,400$ \kms, which
will lead to a very quick separation of the clusters. We should assume that
we are not viewing the cluster system at such a special time when the
projected separation rate reaches more than 4 $h^{-1}$ Mpc per Gyr for a
pair of clusters with only 2.05 $h^{-1}$ Mpc apart.

%\clearpage
\begin{figure} \begin{center}
%\epsscale{0.6}{0.6} \plotone{fig6.eps}
\mbox{\epsfxsize=0.8\textwidth\epsfysize=0.7\textwidth\epsfbox{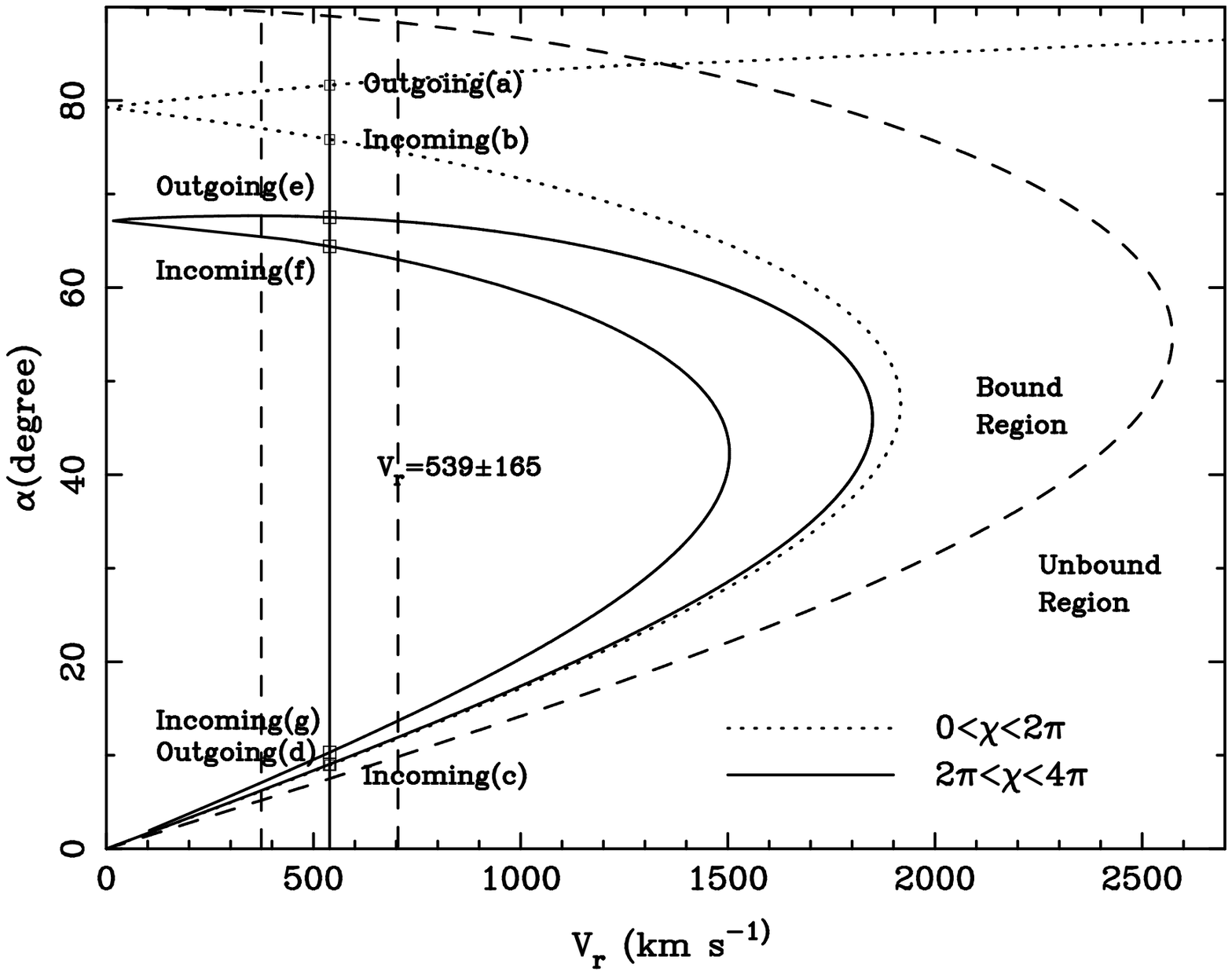}}
\caption{Projection angle $\alpha$ as a function of relative
radial velocity $V_r$ predicted by the two-body model. Four bound
solutions are presented at $V_r=539\pm165$ \kms within $2\pi <
\chi < 4\pi$. The limit between the bound and unbound regions is
also given.}
\end{center} \end{figure}

For cases (d) and (g), the present relative velocity of this bound system
{\em exceeds} the physical velocity dispersion within each cluster, and the
clusters are so close together that they should have begun to coalesce or
have just coalesced. If the system were at such evolutional situations, we
would expect to see some strong merging distortion between these two
clusters in the X-ray surface brightness contours, contrary to the {\it
ROSAT} PSPC image in Fig.~1 of Fabian et al.\ (1997). Therefore these two
solutions can be definitely ruled out.

Cases (e) and (f) are two solutions with larger projection angles, but their
dynamical states are different. Case (e) is an expanding (outgoing) model in
which the last encounter event occured about $2.5h^{-1}$ Gyr ago, and the
cluster pair will expand for another $1.0h^{-1}$ Gyr, reaching a maximum
extent of $5.65 h^{-1}$ Mpc. The true relative velocity is 583 \kms, and
they are $\sim 5.4h^{-1}$ Mpc apart along the direction with a projection
angle of $67^{\circ}.5$. A collapsing model defined in case (f) is also
allowed for this system. According to this model, the two clusters passed
through each other $3.7h^{-1}$ Gyr ago, and reached the last maximum
expansion of $5.0h^{-1}$ Mpc about $0.8h^{-1}$ Gyr ago.

It is rather hard to determine whether the system is collapsing or expanding
at present.  The high resolution observations by the {\it XMM-Newton}
observatory confirmed the enhancement in X-ray flux and temperature in the
region between the two clusters, but no clues of intracluster compression or
shock wave were found (Sakelliou \& Ponman 2004). Gastaldello et al.\ (2003)
pointed out that the profiles of X-ray surface brightness, temperature and
metallicity will shed light on the large-scale dynamics of the binary
cluster system. Sakelliou \& Ponman (2004) presented a clear contour plot of
the residual smoothed images for these two clusters (see figure 6 therein)
where positive residuals can be found on the south-west of the central cD
galaxy in A401 as well as on the north of the one in A399. This indicates
that A401 is moving from south-west to the north-east while A399 is moving
to the south.  This feature seems to favor the scenario in which this pair
of clusters is currently expanding.  Since the projection angles are
$67^{\circ}.5$ in case (e), we can not expect to see a significant
azimuthally asymmetric surface brightness for each cluster in the {\it
ROSAT} PSPC brightness contour maps of this cluster pair (see figure 2 in
Markevitch et al.\ 1998). However, a steeper gradient can be marginally
detectable in the north-east region of A401, and this also points toward an
expanding picture. Therefore case (e) is the more likely solution of the
two-body model.

It should be noted that the two-body model disregards any angular momentum
of the system, and  it ignores the distribution of the matter within
individual clusters which will be important when two clusters are close to
merger. The gravitational interaction from the matter outside the cluster
pair is also neglected. It is well appreciated that dark matter mass
dominate in the overall dynamical mass of individual clusters. Since our
estimate of dynamical mass is based on the virial theorem, two-body model
assumes that the dark matter within a cluster is in approximate virial
equilibrium. Despite the above-mentioned restrictions, two-body model is
still a concise approach which is widely used to discuss the dynamic state
of some gravitationally bound systems.

\section{SUMMARY}

We have investigated the dynamics of the cluster pair A399/A401, using the
existing redshift measurements. We applied the KMM algorithm on a sample of
215 galaxies with known radial velocities, and obtained a robust separation
of this cluster pair. Based on the velocity structure and virial mass
estimate of individual clusters, we explored the two-body model for studying
the merger dynamics between clusters. Because the observational features in
the radio and X-ray wavebands suggest that this pair of clusters have once
experienced a close encounter, we derived four gravitationally bound
solutions within a $\chi$ region between $2\pi$ and $4\pi$. The recent X-ray
data from the {\it ROSAT} and {\it XMM-Newton} observations can be used to
choose the most likely two-body model. In this scenario, the pair of
clusters with a true separation of 5.4$h^{-1}$ Mpc is currently expanding at
583 \kms\ along the direction with a projection angle of $ 67^{\circ}.5$,
and it will reach a maximum extent of $5.65 \,h^{-1}$ Mpc in about
$1.0h^{-1}$ Gyr.

\begin{acknowledgements}
We thank the anonymous referee for his helpful comments. This work is
supported by the National Key Base Sciences Research Foundation under
contract G1999075402, and the National Natural Science Foundation of China
through grant 10273007.
\end{acknowledgements}

%\clearpage

%\vspace{2cm}
%\begin{figure} \begin{center}
%%\epsscale{0.6}{0.6} \plotone{fig6.eps}
%\mbox{\epsfxsize=0.6\textwidth\epsfysize=0.6\textwidth\epsfbox{t-map.eps}}
%\caption{Temperature map of A399 and A401, with
%superposed ROSAT PSPC brightness contour maps.}
%\end{center} \end{figure}

\label{lastpage}

\end{document}